\documentclass[12pt]{iopart}
\usepackage{amsfonts}
\usepackage{amssymb}
\usepackage{graphicx}
\usepackage{bm}
\usepackage{color}

\begin{document}

\title{Two-photon Rabi model: Analytic solutions and spectral collapse}
\author{Liwei Duan$^1$, You-Fei Xie$^1$, Daniel Braak$^2$ and Qing-Hu Chen$^{1,3}$}

\address{$^1$ Department of Physics, Zhejiang University, Hangzhou 310027, China}
\address{$^2$ EP VI and Center for Electronic Correlations and Magnetism, University of Augsburg, 86135 Augsburg, Germany}
\address{$^3$ Collaborative Innovation Center of Advanced Microstructures, Nanjing 210093, China}
\eads{\mailto{daniel.braak@physik.uni-augsburg.de}, \mailto{qhchen@zju.edu.cn}}
\date{\today }

\begin{abstract}
The two-photon quantum Rabi model with quadratic  coupling is studied using
extended squeezed states and we derive $G$-functions for Bargmann index $q=1/4$ and $3/4$.
The simple singularity structure of the $G$-function allows to draw conclusions about
the distribution of eigenvalues along the real axis. The previously found picture of the spectral collapse at critical coupling $g_{\mathrm{c}}$ has to be modified regarding the low lying states, especially the ground state:
We obtain a finite gap between ground state and the continuum of excited states at the collapse point. For large qubit splitting, also other low lying states may be separated from the continuum at $g_{\mathrm{c}}$.
We have carried out a perturbative analysis
allowing for explicit and simple formulae of the eigenstates. Interestingly, a vanishing of the gap between ground state and excited continuum at $g_{\mathrm{c}}$ is obtained in each finite order of approximation. This demonstrates cleary the non-pertubative nature of the excitation gap. We corroborate these findings with a variational calculation for the ground state.
\end{abstract}


\vspace{2pc}

\submitto{\jpa}
\maketitle

\section{Introduction}

The quantum Rabi model (QRM), a major
paradigm for light-matter interaction since its inception in 1936~\cite{Rabi}, has drawn persistent attention due to its applications in
numerous fields ranging from quantum optics to condensed matter physics and even, very recently, to quantum information science.
It describes a two-level system (qubit) coupled linearly to
a cavity electromagnetic mode~\cite{Rabi,Jaynes_Cummings}. The  Hamiltonian can be written as
\begin{equation}
H_{\mathrm{R}}=\frac{\Omega}{2}\sigma _{z}+\omega a^{\dagger }a+g(a^{\dagger }+a)\sigma _{x},
\label{rabimodel}
\end{equation}%
where  $\sigma _{x,z}$ are Pauli matrices describing the two-level system and $a$ ($%
a^{\dagger }$) are the annihilation (creation) bosonic operators of the cavity mode. Although it appears to be much simpler than the hydrogen atom,
it has been considered unsolvable for a long time.  Recently, by using Bargmann-space methods~\cite{Bargmann}, it was shown that this model
is not only exactly solvable but also integrable~\cite{Braak}. The so-called regular spectrum can be obtained by zeros of a  function $G_{\mathrm{R}}(E)$,
i.e. $G_{\mathrm{R}}(E_{n})=0$
entails $E_{n}\in \mathrm{spec} (H_{\mathrm{R}})$.
This $G$-function can be written
explicitly in terms of confluent Heun functions~\cite{Slavyanov}.
 $G_{\mathrm{R}}(E)$  was then recovered within the
extended coherent states approach, which avoids the mapping into the Bargmann space of analytic functions~\cite{Chen2012}.
These results have stimulated extensive research in the QRM
and related models~\cite{twophoton,Moroz,Gardas,Braak2013,Chilingaryan,Maciejewski,Zhong,wang2014,Fanheng,Tomka,Peng,Chennext,Batchelor}%
.

On the other hand, the two-photon QRM has also attracted a lot of attention.
It couples the two-level system to the cavity mode non-linearly and describes a three-level system when the third state can be adiabatically eliminated. It may be realized for Rydberg atoms in microwave
superconducting cavities~\cite{Bertet,Brune} and quantum dots~\cite%
{Stufler,Valle} . The two-photon QRM has also been studied for a long time
both with the RWA~\cite{Puri} and beyond the RWA~\cite%
{Toor,Albert,Zhangyz}. Recently, a realistic implementation of
two-photon quantum Rabi models using trapped ions has been proposed~\cite%
{Felicetti}, which could reach the coupling region corresponding to the interaction-induced spectral collapse. This feature can only be observed in
the deep strong coupling regime of the quantum
Rabi model~\cite{Casanova} and resembles in this respect the well-known superradiant phase transition of the Dicke model~\cite{Wang-Hioe}.

\section{Exact solution using the $\bm{SU(1,1)}$ algebraic structure}

The Hamiltonian of the two-photon QRM is given by
\begin{equation}
H=-\frac{\Omega}{2} \sigma _{x}+a^{\dagger }a+g\left[ \left( a^{\dagger
}\right) ^{2}+a^{2}\right] \sigma _{z},  \label{Hamiltonian}
\end{equation}
where $\Omega $ is the qubit splitting, $a^{\dagger }$ $\left( a\right)$
is the photonic creation (annihilation) operator of the single-mode cavity
with frequency $\omega =1$. We have used the ``spin-boson'' representation~\cite{Duan}, exchanging $\sigma_x$ and $\sigma_z$. The interaction part is quadratic in the boson operators,
while in the original QRM it is linear (see \eref{rabimodel}).

In this section we derive the
G-function found previously for this model~\cite{Chen2012} in a more concise and compact way. First, we
perform a Bogoliubov transformation
\begin{equation}
  b=ua+va^{\dagger },\qquad b^{\dagger }=ua^{\dagger }+va,
  \label{bogol}
\end{equation}
to a new bosonic operators. With
\begin{equation}
u=\sqrt{\frac{1+\beta }{2\beta }},\qquad v=\sqrt{\frac{1-\beta }{2\beta }},
\end{equation}
and $\beta =\sqrt{1-4g^2}$, the upper diagonal matrix element of the
Hamiltonian becomes
\begin{equation*}
H_{11}=a^{\dagger }a+g\left[ \left( a^{\dagger }\right) ^2+a^2\right] =\frac{%
b^{\dagger }b-v^2}{u^2+v^2}.
\end{equation*}
In terms of $b, b^\dagger$, the Hamiltonian reads
\begin{equation}
H=\left(
\begin{array}{ll}
\frac{b^{\dagger }b-v^2}{u^2+v^2} & ~-\frac \Omega 2 \\
~~-\frac \Omega 2 & \;\;\;H_{22}%
\end{array}
\right) ,
\end{equation}
with
\begin{equation*}
H_{22}=\left( u^2+v^2+4guv\right) b^{\dagger }b-2uv\left[ \left( b^{\dagger
}\right) ^2+b^2\right] +2guv+v^2.
\end{equation*}
The operators $ b^\dagger b$, $(b^\dagger)^2$, $b^2$ provide a representation
of the non-compact Lie algebra $su(1,1)$:
With
\begin{equation*}
K_0=\frac 12\left( b^{\dagger }b+\frac 12\right),\qquad K_{+}=\frac 12\left(
b^{\dagger }\right) ^2,\qquad K_{-}=\frac 12b^2,
\end{equation*}
we have
\begin{eqnarray*}
\left[ K_0,K_{\pm }\right] = \pm K_{\pm }, \qquad \left[ K_{+},K_{-}\right] = -2K_0.
\end{eqnarray*}
The quadratic Casimir operator $C$ of the algebra is given by
\begin{equation*}
C=K_{+}K_{-}-K_0\left( K_0-1\right).
\end{equation*}
The infinite-dimensional unitary representations of $su(1,1)$ are labeled by the value $q$ of $C$, the Bargmann index. Here, the Hilbert space $\cal H$ generated by $b^\dagger$ on the
state $|0\rangle_{\mathrm{b}}$ anihilated by $b$, separates into two $H$-invariant subspaces,
${\cal H}={\cal H}_{\frac 14}\oplus{\cal H}_{\frac 34}$ for $q=\frac 14,\frac 34$. A basis of
${\cal H}_q$ is given by the normalized states
\begin{eqnarray}
\left| q,n\right\rangle_{\mathrm{b}} =\frac{\left( b^{\dagger }\right) ^{2\left(
n+q-\frac 14\right) }}{\sqrt{\left[ 2\left( n+q-\frac 14\right) \right] !}}%
\left| 0\right\rangle _{\mathrm{b}}=\left| 2\left( n+q-\frac 14\right) \right\rangle _{\mathrm{b}},
\\
q =\frac 14,\frac 34,\qquad n=0,1,2,...\infty.\nonumber
\end{eqnarray}
The operators
satisfy
\begin{eqnarray*}
K_{+}\left| q,n\right\rangle _{\mathrm{b}} =\sqrt{\left( n+q+\frac 34\right) \left(
n+q+\frac 14\right)}\left| q,n+1\right\rangle _{\mathrm{b}}, \\
K_{-}\left| q,n\right\rangle _{\mathrm{b}} =\sqrt{\left( n+q-\frac 14\right) \left(
n+q-\frac 34\right)}\left| q,n-1\right\rangle _{\mathrm{b}}, \\
K_0\left| q,n\right\rangle _{\mathrm{b}} =\left( n+q\right) \left| q,n\right\rangle _{\mathrm{b}}.
\end{eqnarray*}
Note that the vacuum with respect to the original boson operators $a,a^\dagger$,
$|0\rangle_{\mathrm{a}}$, with the property $a|0\rangle_{\mathrm{a}}=0$, may be expressed
in terms of $\left\vert \frac 14,n\right\rangle _{\mathrm{b}}$ as
\begin{equation*}
\left\vert 0\right\rangle _{\mathrm{a}}=\sum_{n=0}^{\infty }z_{n}^{(\frac{%
1}{4})}\left\vert \frac{1}{4},n\right\rangle _{\mathrm{b}},
\end{equation*}%
because the decomposition ${\cal H}={\cal H}_{\frac 14}\oplus{\cal H}_{\frac 34}$ is left invariant by the Bogoliubov transformation \eref{bogol}.
We can write therefore $|0\rangle_{\mathrm{a}}=\vert \frac 14,0\rangle_{\mathrm{a}}$.
The condition $a\left\vert 0\right\rangle _{\mathrm{a}}=0,$ leads to
\begin{equation}
z_{n}^{(\frac{1}{4})}\varpropto \frac{\sqrt{\left( 2n\right) !}}{n!}\left(
\frac{v}{2u}\right) ^{n}.  \label{q1_c}
\end{equation}%
The lowest lying state (with respect to the $a$-operators) in  ${\cal H}_{\frac 34}$ reads then
\begin{eqnarray*}
\left\vert \frac{3}{4},0\right\rangle _{\mathrm{a}} =a^{\dagger }\left\vert \frac 14,0\right\rangle _{\mathrm{a}}=\left( ub^{\dagger }-vb\right) \sum_{n=0}^{\infty
}z_{n}^{(\frac{1}{4})}\left\vert \frac{1}{4},n\right\rangle _{\mathrm{b}}
=\sum_{n=0}^{\infty }z_{n}^{(\frac{3}{4})}\left\vert \frac{3}{4}%
,n\right\rangle _{\mathrm{b}},
\end{eqnarray*}%
where
\begin{equation}
z_{n}^{(\frac{3}{4})}\varpropto \frac{\sqrt{\left( 2n+1\right) !}}{n!}\left(
\frac{v}{2u}\right) ^{n}.  \label{q2_c}
\end{equation}%
In summary,
\begin{equation}
z_{n}^{(q)}\varpropto \frac{\sqrt{\left[ 2\left( n+q-\frac{1}{4}\right) %
\right] !}}{n!}\left( \frac{v}{2u}\right) ^{n}  \label{q12_c}.
\end{equation}

In terms of the $K_0$, $K_\pm$,
the Hamiltonian reads
\begin{equation}
H=\left(
\begin{array}{ll}
\frac{\left( 2K_0-\frac 12\right) -v^2}{u^2+v^2} & ~-\frac \Omega 2 \\
~~-\frac \Omega 2 & \;\;\;H_{22}^{\prime }%
\end{array}
\right) ,
\end{equation}
where
\begin{equation*}
H_{22}^{\prime }=\left( u^2+v^2+4guv\right) \left( 2K_0-\frac 12\right)
-4uv\left( K_{+}+K_{-}\right) +2guv+v^2.
\end{equation*}

An eigenfunction $|\psi,E\rangle$ of $H$ with eigenvalue $E$ may be expanded in terms of the $b$-operators as
\begin{equation}
\left\vert \psi,E\right\rangle =\left( \
\begin{array}{l}
\sum_{m=0}^{\infty }\sqrt{\left[ 2\left( m+q-\frac{1}{4}\right) \right] !}%
e_{m}^{(q)}\left\vert q,m\right\rangle _{\mathrm{b}} \\
\sum_{m=0}^{\infty }\sqrt{\left[ 2\left( m+q-\frac{1}{4}\right) \right] !}%
f_{m}^{(q)}\left\vert q,m\right\rangle _{\mathrm{b}}%
\end{array}%
\right) ,
\end{equation}%
Projecting both sides of the
Schr\"odinger equation onto $_{\mathrm{b}}\left\langle q,n\right\vert $
gives a linear relation between
coefficients $e_{n}^{(q)}$ and $f_{n}^{(q)}$,%
\begin{equation}
e_{n}^{(q)}=\frac{\frac{\Omega }{2}}{\frac{2\left( n+q-\frac{1}{4}\right)
-v^{2}}{u^{2}+v^{2}}-E}f_{n}^{(q)},  \label{1-1}
\end{equation}%
and
\begin{eqnarray}
\fl 8uv\left( n+q+\frac{3}{4}\right) \left( n+q+\frac{1}{4}\right)
f_{n+1}^{(q)}=-2uvf_{n-1}^{(q)}\\
+\left[ \left( u^{2}+v^{2}+4guv\right) \left( 2\left( n+q\right) -\frac{1}{2}%
\right) +2guv+v^{2}-E\right] f_{n}^{(q)}
-\frac{\Omega }{2} e_{n}^{(q)}
.\nonumber
\end{eqnarray}
We obtain a three-term recurrence relation
\begin{eqnarray}
\fl f_{n+1}^{(q)}=\frac{\left( 1+4g^{2}\right) \left( n+q\right) -\beta
^{2}\left( x+q\right) -\frac{\Omega ^{2}}{16\left( n-x\right) }}{4g\left(
n+q+\frac{3}{4}\right) \left( n+q+\frac{1}{4}\right) }f_{n}^{(q)}-\frac{%
f_{n-1}^{(q)}}{4\left( n+q+\frac{3}{4}\right) \left( n+q+\frac{1}{4}\right) },
\label{recure_2p}
\end{eqnarray}%
where $x=\frac{E}{2\beta}+\frac{v^2}{2}-q+\frac{1}{4}$, the coefficients $f_{n}^{(q)}$
are calculated with initial conditions $f_{0}^{(q)}=1$, $f_{-1}^{(q)}=0$.

Because of parity invariance in each space ${\cal H}_q$~\cite{Chen2012}, projecting the wavefunction $|\psi,E\rangle$ onto $|\uparrow \rangle|q,0\rangle _{\mathrm{a}}$ and $|\downarrow \rangle |q,0\rangle _{\mathrm{a}}$ respectively, we can define the two-photon
G-function as
\begin{equation}
G_{\pm }^{(q)}(x)=\sum_{n=0}^{\infty }f_{n}^{(q)}\;\left[ 1+\Pi \frac{\Omega
}{4\beta \left( n-x\right) }\;\right] \frac{\left[ 2\left( n+q-\frac{1}{4}%
\right) \right] !}{n!}\left( \frac{v}{2u}\right) ^{n},  \label{G-2photon}
\end{equation}%
where $\Pi =\pm 1$, corresponding to positive(negative) parity. So far, we
have just re-derived the $G$-function in a more compact and concise way compared to \cite{Chen2012}

\begin{figure}[tbp]
\includegraphics[width=12cm]{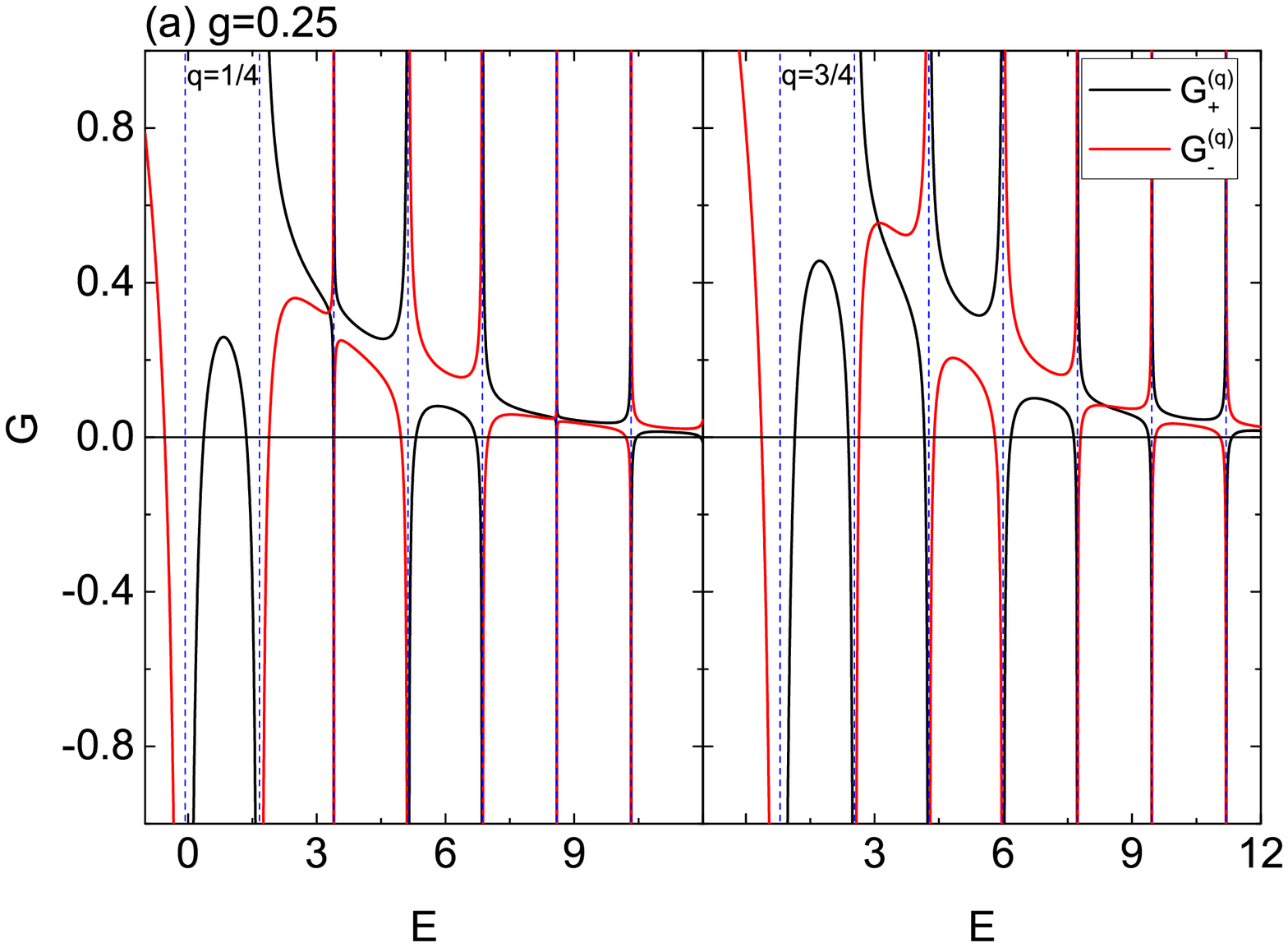} %
\includegraphics[width=12cm]{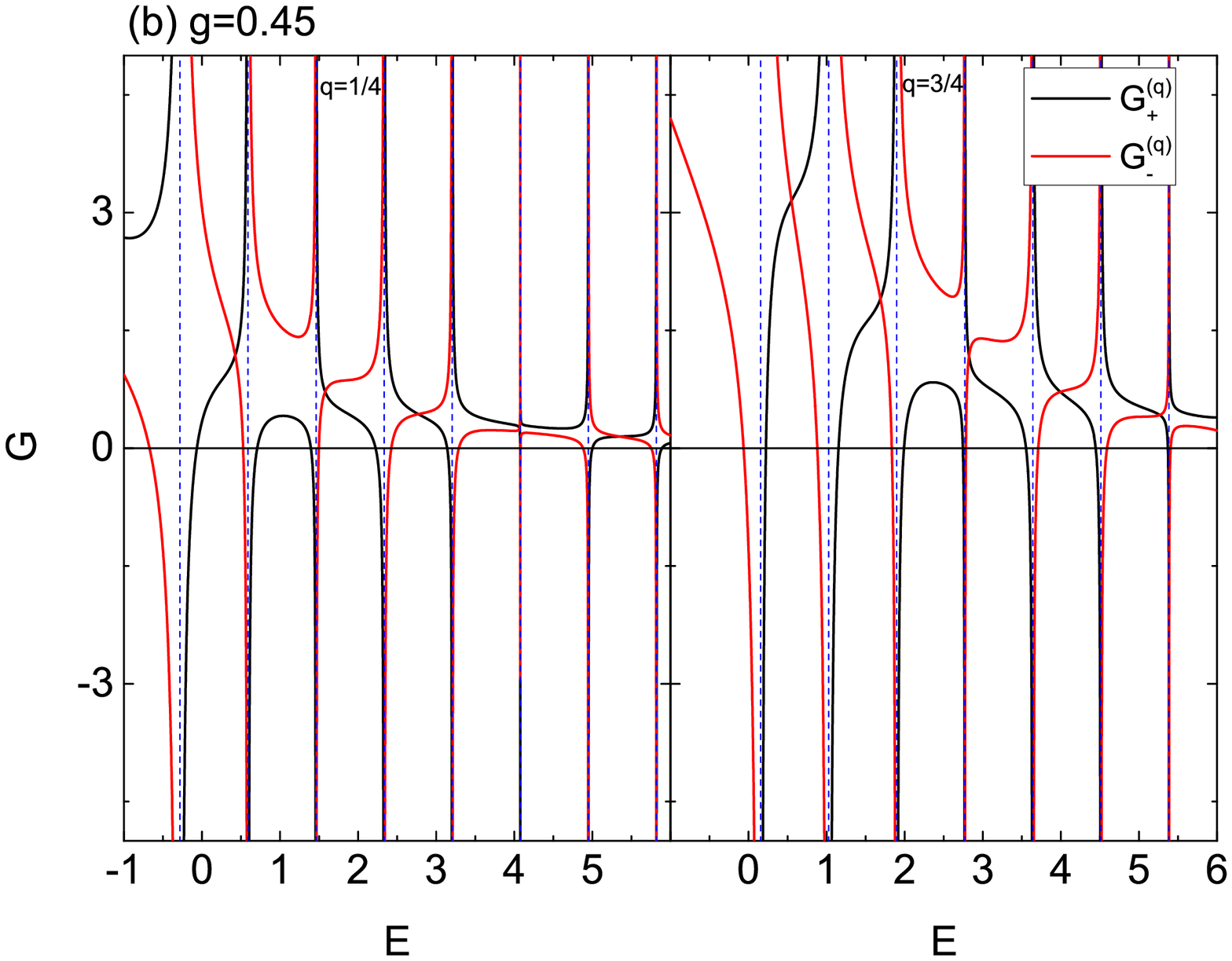}
\caption{(Color online) G curves for the two-photon QRM at $\Omega =1$, (a) $%
g=0.25$ (upper) and (b)  $g=0.45$ (lower). }
\label{gfunction_2p}
\end{figure}

We plot the two-photon $G$-function in \fref{gfunction_2p} for $\Omega =1$, two values of $g$, $0.25$ and $0.45$, and both Bargmann indices
$q=\frac{1}{4}$ and
$\frac{3}{4}$. The zeros give the location of the energy spectrum, which
is plotted in \fref{spectrum_tp} as function of $g$.

The poles in \fref{gfunction_2p} correspond to values of $E=\beta \left( 2(n+q-
\frac{1}{4})-v^{2}\right) =2\beta (n+q)-\frac{1}{2}$. The position of the first pole corresponds to $E=2\beta q-\frac{1}{2}$.
The distance between adjacent poles is $2\beta $
and vanishes as $g\rightarrow \frac{1}{2}=g_{\mathrm{c}}$. Therefore zeros of $G_\pm^{(q)}(E)$ (energy levels)
between
two poles will collapse towards $-\frac{1}{2}$ when $g\rightarrow \frac{1}{2}$, as shown in \fref{spectrum_tp}.

\section{Spectral collapse and energy gap}{\label{exact}}

The  energy of the ground state does not tend to $-\frac{1}{2}$ when $%
g\rightarrow \frac{1}{2}$, in sharp contrast with almost all of the excited states. This is seen in all previous numerical calculations of the spectra, but has never been discussed in detail. Here, we will present an explanation with help of the
analytical exact solutions.

The first pole ($n=0$) of $G_\pm^{(q)}(E)$ forms the zeroth baseline in the spectral graph,
\begin{equation}
E_{\mathrm{pol}}^{(1)}(g)=2\sqrt{1-4g^{2}}q-\frac{1}{2}  \label{pol_1}
\end{equation}%
and approaches $2q-\frac{1}{2}$ in the weak coupling limit $g\rightarrow 0$.
On the other hand, for $g=0$, the qubit is decoupled from the cavity, all
eigenenergies are easily obtained as
\begin{equation}
E^n_{\Pi }(g=0)=\Pi \frac{\Omega }{2}\left( -1\right) ^{n}+2\left( n+q-\frac{1%
}{4}\right)  \label{g_0}
\end{equation}%
where $n=0,1,2,3...$. If $E^n_{\Pi }(g=0)<E_{\mathrm{pol}}^{(1)}(g=0)$, i.e.%
\begin{equation}
n<-\Pi \frac{\Omega }{4}(-1)^{n},\qquad n=0,1,2,...  \label{criterion}
\end{equation}%
then the energy level $E^n_\Pi(g)$ will be smaller than $E^{(1)}_{\mathrm{pol}}(g)$ for $g>0$ as well until an exceptional solution~\cite{Braak} is reached where the pole coincides with an energy eigenvalue and the zeroth baseline is crossed. This is possible if the exceptional solution is not Juddian (doubly-degenerate)~\cite{Duan,notjud, Braak15}. We shall see that also the opposite occurs: an energy level above the first pole for small $g$ crosses the zeroth baseline at $g_0$ and lies below the first pole for $g>g_0$.
These levels correspond to zeros of the $G-$function which are not pinched between the poles as $g$ approaches $g_{\mathrm{c}}=\frac 12$ and will therefore not collapse into the continuum.

\begin{figure}[tbp]
\includegraphics[width=8cm]{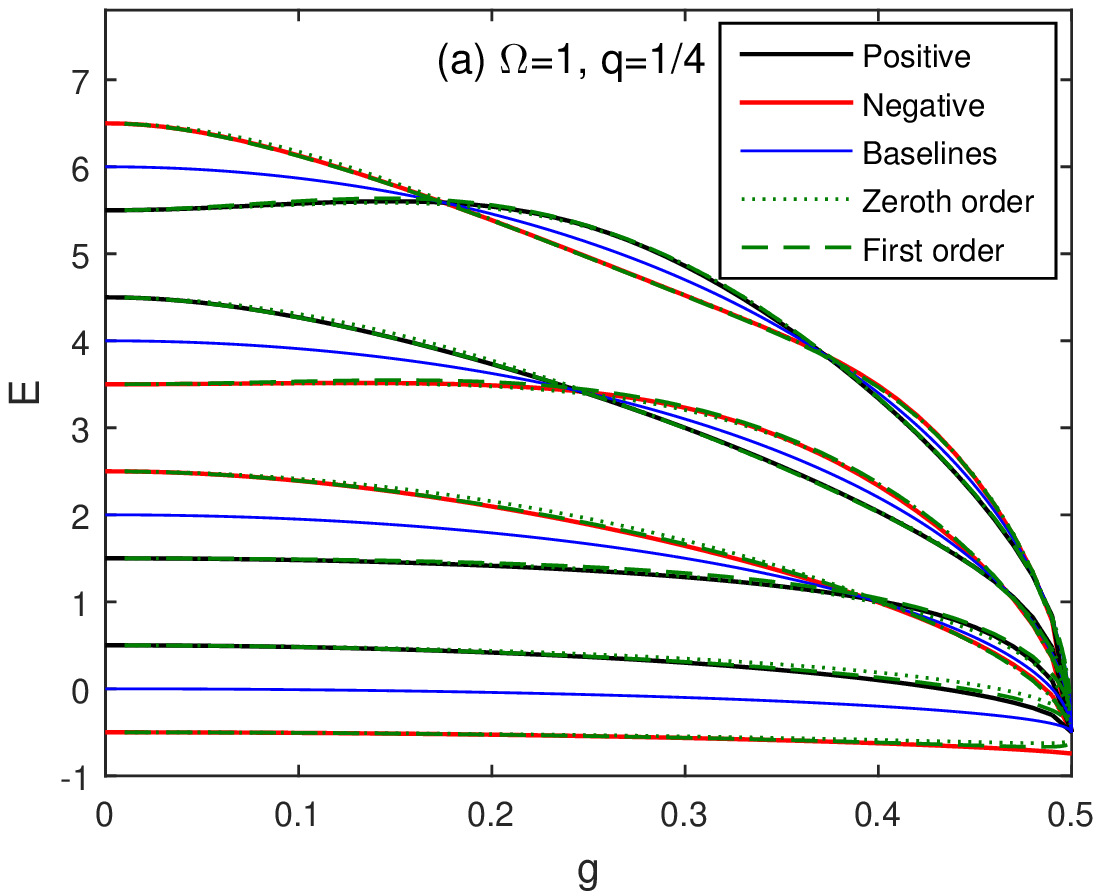} %
\includegraphics[width=8cm]{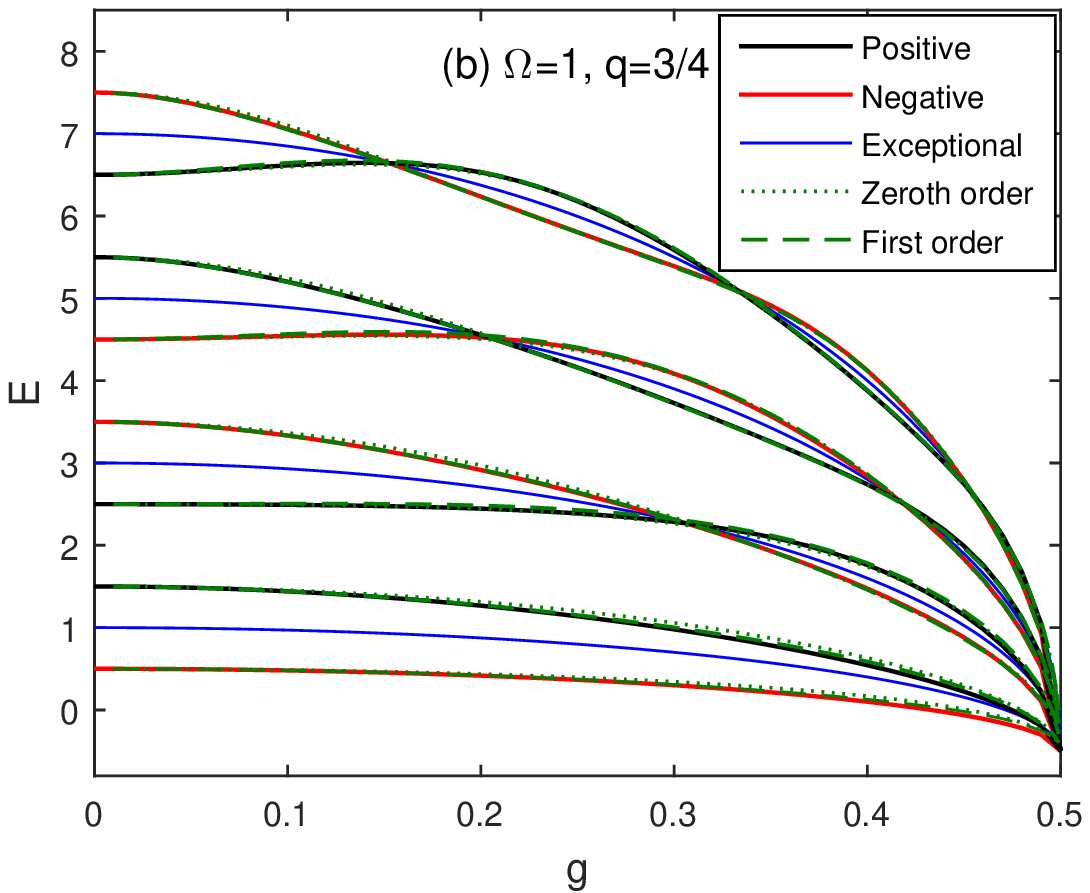} %
\includegraphics[width=8cm]{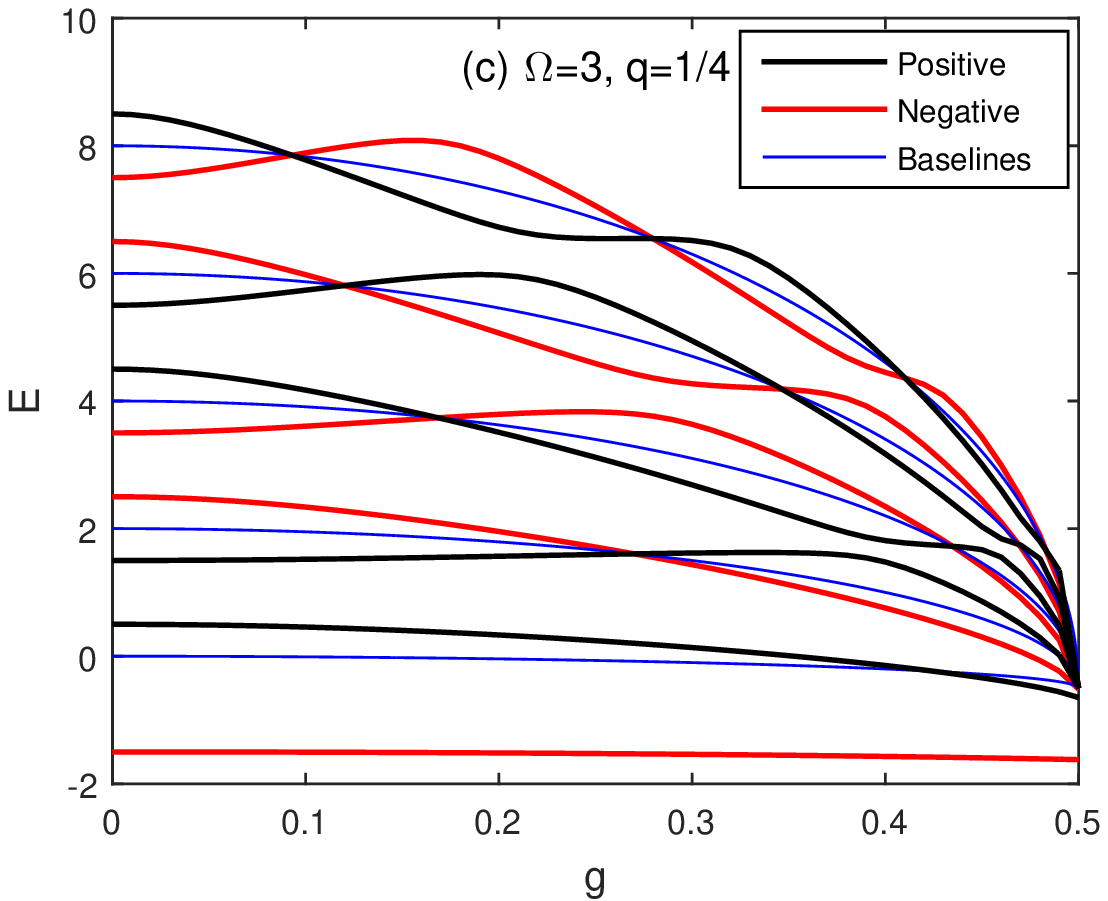} %
\includegraphics[width=8cm]{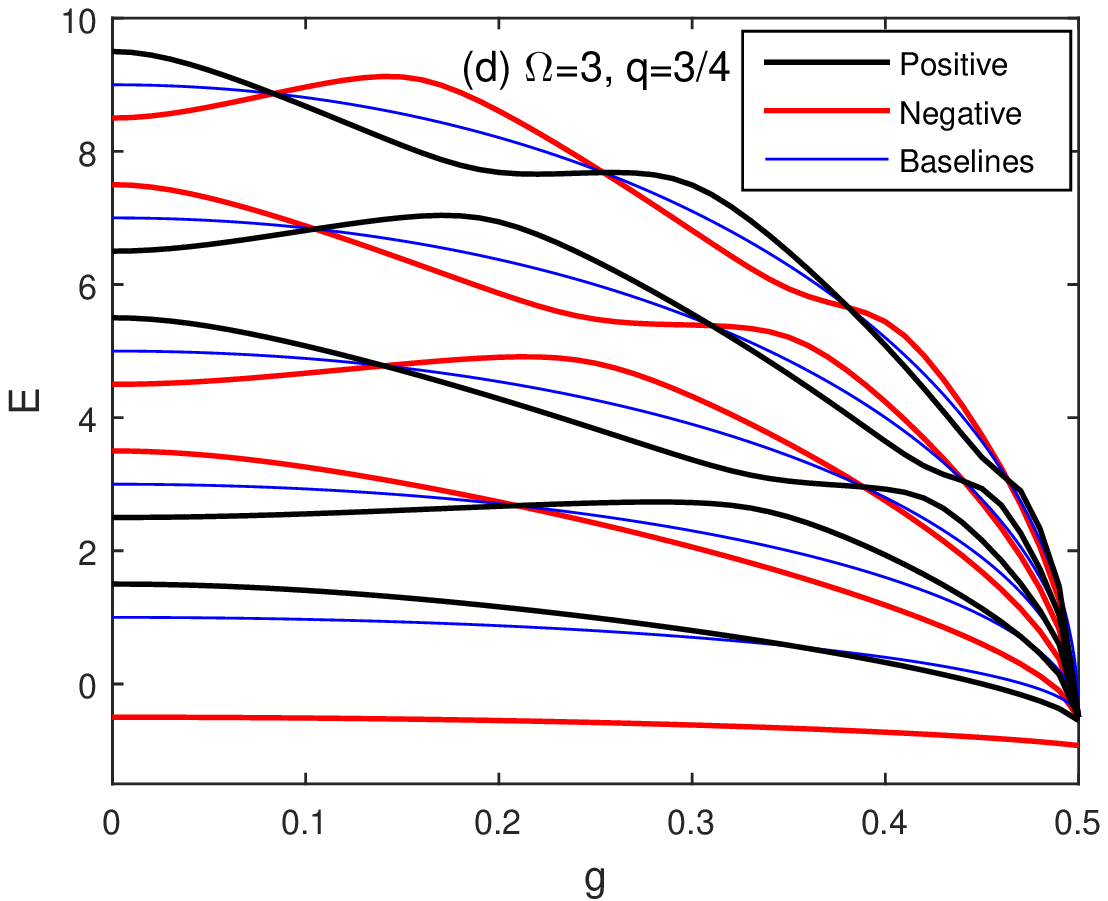}
\caption{(Color online) Energy spectra obtained from G-function for $%
\Omega=1 $ (upper panel) and $\Omega=3$ (low panel), $q=1/4$ ( left panel)
and $q=3/4$ (right panel). The baselines (blue solid) are also
presented. The zeroth order approximation (dotted) and first order
approximation (dashed) are given only for $\Omega=1$ (upper panel). }
\label{spectrum_tp}
\end{figure}

The ground state belongs to negative parity for each $q$ and does not cross the zeroth
baseline for the examples in \fref{spectrum_tp}. It will always be separated by a finite gap from the continuum at $g=\frac{1}{2}$. For large $\Omega=3$, we have in the lower panels of \fref{spectrum_tp} an example of an excited state with positive parity crossing the zeroth baseline at $g_0<g_{\mathrm{c}}$. This state will also not collapse as $g=\frac 12$ is reached.
Because several eigenstates for each parity are located below the first pole for small $g$ according to \ref{criterion}, and correspond to zeros of the $G$-function in a pole-free region, none of them is constrained by the argument above and may be separated from the continuum at the critical coupling, if they do not cross the zeroth baseline for some $g<g_{\mathrm{c}}$.

\section{Finite-dimensional approximations}{\label{approx}}

Similar to  $b$ and $b^{\dag}$, we can introduce another
set of operators
\begin{equation}
c=ua-va^{\dagger },\qquad c^{\dagger }=ua^{\dagger }-va,
\end{equation}
which removes the terms $a^2$, $(a^\dag)^2$ in the other diagonal element of the Hamiltonian matrix
element
\begin{equation*}
H_{22}=a^{\dagger }a-g\left[ \left( a^{\dagger }\right) ^2+a^2\right] =\frac{%
c^{\dagger }c-v^2}{u^2+v^2}.
\end{equation*}

The corresponding basis reads
\begin{eqnarray}
\left| q,n\right\rangle_{\mathrm{c}} =\frac{\left( c^{\dagger }\right) ^{2\left(
n+q-\frac 14\right) }}{\sqrt{\left[ 2\left( n+q-\frac 14\right) \right] !}}%
\left| 0\right\rangle _{\mathrm{c}}=\left| 2\left( n+q-\frac 14\right) \right\rangle _{\mathrm{c}},
\\
q =\frac 14,\frac 34,\qquad n=0,1,2,....\infty,\nonumber
\end{eqnarray}

The Hamiltonian reads in terms of $b$
and $c$ operators
\begin{equation}
H=\left(
\begin{array}{ll}
\beta \left( b^{\dagger }b-v^{2}\right) & ~-\frac{\Omega }{2} \\
~~-\frac{\Omega }{2} & \;\beta \left( c^{\dagger }c-v^{2}\right)%
\end{array}%
\right) ,
\end{equation}%
For each parity $\Pi$ we write the wavefunction as
\begin{equation}
\left\vert\psi,E\right\rangle _{q}=\left( \
\begin{array}{l}
\sum_{n=0}^{\infty }u_{n}^{(q)}\left\vert q,n\right\rangle _{\mathrm{b}} \\
-\Pi \sum_{n=0}^{\infty }\left( -1\right) ^{n}u_{n}^{(q)}\left\vert
q,n\right\rangle _{\mathrm{c}}%
\end{array}%
\right) .
\end{equation}%
From the Schr\"{o}dinger equation it follows then
\begin{equation*}
\beta \left( b^{\dagger }b-v^{2}\right) \sum_{n=0}^{\infty
}u_{n}^{(q)}\left\vert q,n\right\rangle _{\mathrm{b}}+\Pi \frac{\Omega }{2}%
\sum_{n=0}^{\infty }\left( -1\right) ^{n}u_{n}^{(q)}\left\vert
q,n\right\rangle _{\mathrm{c}}=E\sum_{n=0}^{\infty }u_{n}^{(q)}\left\vert
q,n\right\rangle _{\mathrm{b}}.
\end{equation*}%
Projection on $\left\vert q,m\right\rangle _{\mathrm{b}}$ gives%
\begin{equation}
\beta \left( 2(m+q-\frac{1}{4})-v^{2}\right) u_{m}^{(q)}+\Pi
\sum_{n=0}^{\infty }u_{n}^{(q)}\ D_{m,n}^{(q)}=Eu_{m}^{(q)}  \label{main_ana}
\end{equation}%
with%
\begin{eqnarray}
D_{m,n}^{(q)}&=\frac{\Omega }{2}\left( -1\right) ^{n}\ _{\mathrm{b}}\left\langle
q,m\right\vert \left. q,n\right\rangle _{\mathrm{c}}\nonumber\\
&=\frac{\Omega }{2}\ \left(
-1\right) ^{m}\beta ^{\frac{1}{2}}\sqrt{\frac{\left[ 2\left( n+q-\frac{1}{4}%
\right) \right] !}{\left[ 2\left( m+q-\frac{1}{4}\right) \right] !}}%
P_{m+n+2\left( q-\frac{1}{4}\right) }^{m-n}\left( \beta \right)  \label{D_mn}
\end{eqnarray}%
where $P_{m+n}^{m-n}\left( \beta \right) $ is an associated Legendre
polynomial, which is defined for all values of integer $\ m$ and $n$.
Obviously, when $g\rightarrow \frac 12$, $D_{mn}\rightarrow 0,$ which will be
used later.

We can use the set of equations from \eref{main_ana} to
diagonalize the Hamiltonian and get numerical exact solutions with some truncation in $n$.
We define the $N$%
-th order approximation by selecting $N$
coefficients $u_{n}^{(q)}$, $\left( n=m,m+1,...,m+N\right) $ in the \eref{main_ana} and neglect the other terms.

In zeroth order, we set $N=0$ and have
\begin{equation}
\beta \left( 2(m+q-\frac{1}{4})-v^{2}\right) u_{m}^{(q)}+\Pi u_{m}^{(q)}\
D_{m,m}^{(q)}=Eu_{m}^{(q)}
\end{equation}%
which gives the eigenenergy immediately%
\begin{equation}
E_{m}^{(0)}\left( \Pi \right) =\beta \left( 2(m+q-\frac{1}{4})-v^{2}\right)
+\Pi D_{m,m}^{(q)}  \label{ana_zero}
\end{equation}%
The ground-state energy is $E_{0}^{(0)}\left( \Pi =-1\right) $, of negative
parity.

For the $N=1$ we obtain an explicit
expression for the energy as well. For the excited states, we  have two
equations for two coefficients
\begin{eqnarray}
\left[ \beta \left( 2(m+q-\frac{1}{4})-v^{2}\right) +\Pi D_{m,m}^{(q)}\right]
u_{m}^{(q)}+\Pi u_{m+1}^{(q)}\ D_{m,m+1}^{(q)} =Eu_{m}^{(q)},  \label{1st_a}
\\
\Pi D_{m+1,m}^{(q)}u_{m}^{(q)}+\left[ \beta \left( 2(m+q+\frac{3}{4}%
)-v^{2}\right) +\Pi D_{m+1,m+1}^{(q)}\right] u_{m+1}^{(q)} =Eu_{m+1}^{(q)}.
\label{1st_b}
\end{eqnarray}%
Obviously, for each $m=0,1,2,...$, we have four solutions from the above
equation, two of them are redundant. At weak coupling, the parity for each eigenstate is fixed: even $m$ for positive parity and odd $m$ for negative parity.
It follows $\Pi \left( -1\right) ^{m}=1$. Therefore, we may replace $\Pi$ by $(-1)^m$ and obtain for the
eigenenergies of the excited states
\begin{eqnarray}
\fl E_{m}^{(1)} =\beta \left( 2(m+q+\frac{1}{4})-v^{2}\right) +\frac{(-1)^{m}}{%
2}(D_{m,m}+D_{m+1,m+1})  \nonumber \\
\pm \frac{1}{2}\sqrt{\left[ (-1)^{m}(D_{m,m}-D_{m+1,m+1})-2\beta \right]
^{2}+4D_{m,m+1}^{2}},m=0,1,2,...  \label{ana_first}
\end{eqnarray}%
Note that for each $m$, we have two solutions with the same parity.

The ground-state energy in the first order approximation is given by
\eref{1st_a} and \eref{1st_b} for$\ m=0,q=\frac{1}{4}$, and $\Pi =-1$
\begin{eqnarray}
\fl E_{\mathrm{GS}}^{(1)} =-\frac{1}{2}+\frac{3}{2}\beta -\frac{1}{2}\left( \ D_{0,0}^{(%
\frac{1}{4})}+\ D_{1,1}^{(\frac{1}{4})}\right)-\frac{1}{2}\sqrt{\left( D_{1,1}^{(\frac{1}{4})}-\ D_{0,0}^{(\frac{1}{4}%
)}-2\beta \right) ^{2}+4D_{1,0}^{(\frac{1}{4})}\ D_{0,1}^{(\frac{1}{4})}}.
\label{GS_first}
\end{eqnarray}

The arbitrary $N$-th order approximation can be performed straightforwardly.
There are $2\left( N+1\right) $ solutions for each value of $m$.

The energy levels in zeroth order and first order approximation are also
presented in \fref{spectrum_tp} ($\Omega =1$, upper level). The
energy levels agree even  in the zero order approximation quite well with the
exact ones. Interestingly, the spectral collapse is exhibited already in
zeroth order. This is not strange, because the matrix
elements $D_{mn}$ in \eref{main_ana} tends to zero if $g$ approaches $
\frac 12$. Actually in any order of the analytic approximation, the energy
for all eigenstates  approaches  $-\frac{1}{2}$, including the ground
state, leading to a complete collapse. This is not true, as shown in
section~\ref{exact}. The finite order approximations break down for all $N$
when the critical point $g=\frac 12$ is approached. In the following section, we  perform a variational analysis of the ground state to show this in an alternative way.

\begin{figure}[tbp]
\includegraphics[width=14cm]{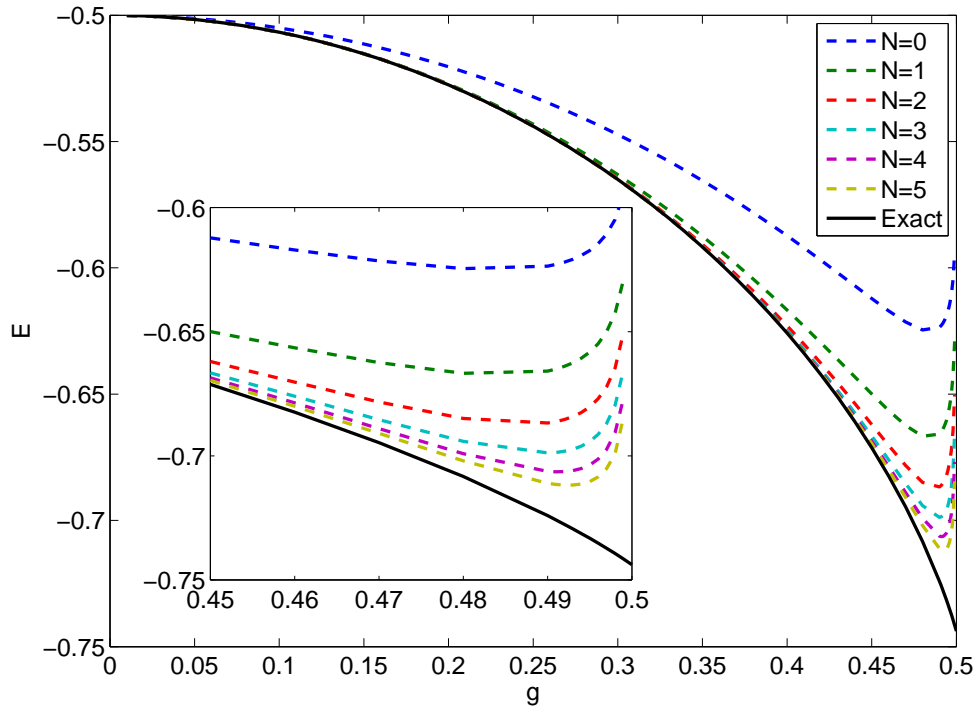}
\caption{(Color online) Comparison of the ground state energies as
function of $g$ calculated by the $N$-th order approximation with the exact
results at $\Omega =1$. }
\label{GS}
\end{figure}

\section{Variational calculation for the ground state}{\label{ground}}

\bigskip

The ground state corresponds to $q=\frac{1}{4}$ and negative parity, so
we have%
\begin{equation}
\beta \left( 2m-v^{2}\right) u_{m}^{(q)}-\sum_{n=0}^{N_{\mathrm{tr}}}u_{n}^{(q)}\
D_{m,n}^{(q)}=Eu_{m}^{(q)},  \label{GS_energy}
\end{equation}%
where $N_{\mathrm{tr}}$ is truncation number. The lowest energy obtained from
the above eigenvalue problem will give the ground-state energy $%
E_{\mathrm{GS}}^{(N_{\mathrm{tr}})}$ in the $N_{\mathrm{tr}}$-th order of approximation. $E_{\mathrm{GS}}^{(1)}$%
corresponds to $N_{\mathrm{tr}}=1$.

We plot the GS energy as a function of coupling strength in different order
of approximations for $\Omega =1$, in \fref{GS}. The GS energy becomes
closer to the exact one as the approximation order increases, but it tends
to the collapse value $-\frac{1}{2}$ finally as $g$ goes to $\frac 12$ in any finite order
approximation.
This holds as well for all states below the continuum discussed in section ~\ref{exact}.
It indicates that any finite order approximation misses the low energy features of the spectrum  at the critical coupling.

We elucidate this finding by performing a variational study for the ground-state with negative parity. The trial wavefunction reads for $\Pi =-1,q=\frac{1}{4}$
\begin{equation}
\left\vert {GS_{trial}}\right\rangle _{q=\frac{1}{4}}\varpropto \left( \
\begin{array}{l}
\left\vert \frac{1}{4},0\right\rangle _{b^{\prime }} \\
\left\vert \frac{1}{4},0\right\rangle _{c^{\prime }}
\end{array}
\right),
\end{equation}
with
\begin{equation}
b=u^{\prime }a+v^{\prime }a^{\dagger },\qquad c=u^{\prime }a-v^{\prime }a^{\dagger },
\end{equation}%
where
\begin{equation}
u^{\prime }=\cosh r,\qquad v^{\prime }=\sinh r,
\end{equation}%
and the variational parameter $r$. The corresponding energy reads then
\begin{equation}
E(r)=-\frac{\Omega }{2}\left[ 1-\tanh ^{2}\left( 2r\right) \right]
^{1/4}+\omega \sinh ^{2}\left( r\right) -g\sinh \left( 2r\right).
\end{equation}%
Minimizing $E(r)$ with respect to $r$ gives a variational estimate for the ground-state energy.

\begin{figure}[tbp]
\includegraphics[width=14cm]{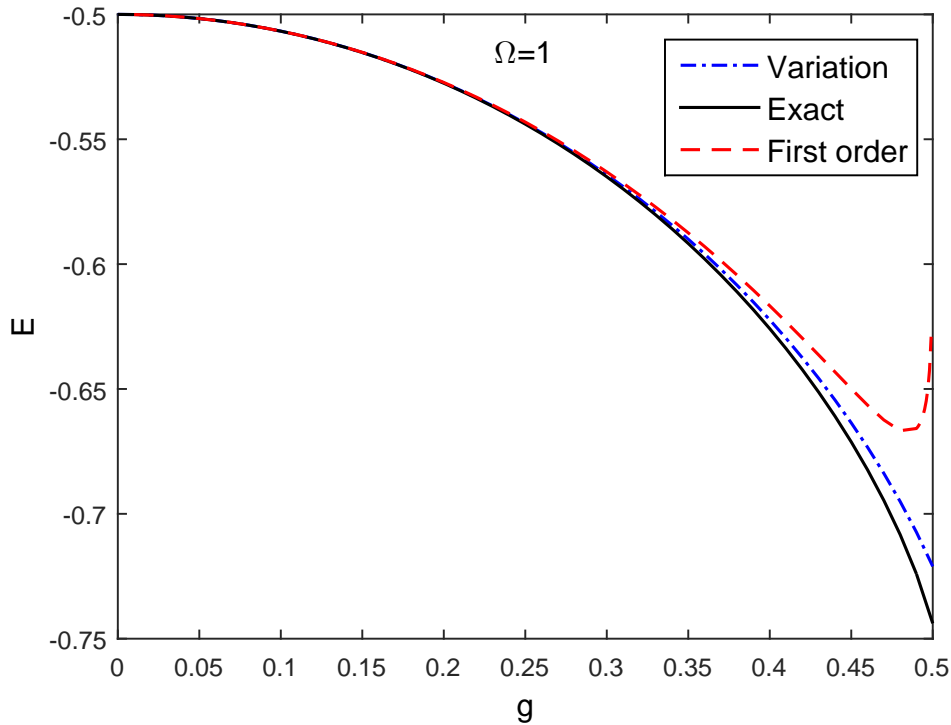}
\caption{(Color online) Comparison of the ground state energies as
function of $g$ calculated by the variational ansatz, the first order
approximation and the exact results at $\Omega =1$. }
\label{C_GS}
\end{figure}

In \fref{C_GS}, we compare the GS energies obtained by the above
variational ansatz with those of the first-order approximation and the exact
G-function technique. It is found that the variational GS energy is much better
than obtained in the first-order approximation. More interestingly, the
variational GS energy does not collapse towards $-\frac 12$. This proves that the lower edge of the continuum cannot coincide with the groundstate of the system which is always gapped.

\section{Conclusions}

In this work, we have derived the $G$-function for the two-photon QRM in a
concise and compact way, by using extended squeezed states for each Bargmann index. Zeros of
the $G$-function determine the regular spectrum. The average distance between energy levels is dictated
by the pole structure of $G(E)$. If
the $n$-th level for any finite $n$ is located between two poles as $g$ tends to $\frac 12$, this level will
collapse to the value $-\frac 12$ at $g=\frac 12$.
The ground state is located below the first pole for $g \ll 1$ and remains so until $g_{\mathrm{c}}$ is reached. However, a crossing  of the zeroth baseline from below cannot be ruled out, because non-degenerate exceptional solutions are possible for $n=0$. It seems that these always belong to excited states with positive parity and large $\Omega$: the zeroth baseline is crossed from above so that this state lies in the gap between ground state and continuum at $g=g_{\mathrm{c}}$.
In general the $G$-function has several zeros below $E=-\frac 12$ for large $\Omega$ and small $g$. All of them seem to remain separated from the continuum at the collapse point.
We have calculated explicit solutions in a finite-dimensional approximation
scheme and found that the zeroth order describes the
collapse well but the gap and the discrete levels inside do not appear in any finite order approximation. Nevertheless, the existence of the gap itself can be proven by a simple variational analysis of the ground state.

\ack{ This work was supported by National Natural
Science Foundation of China under Grant Nos. 11174254 and 11474256. D.B.
acknowledges support by TRR80 of the Deutsche Forschungsgemeinschaft.}

\end{document}